\documentstyle[aps,preprint,tighten,epsfig]{revtex}
\def\beq{\begin{equation}}
\def\eeq{\end{equation}}
\def\beqa{\begin{eqnarray}}
\def\eeqa{\end{eqnarray}}
\def\MeV{\nobreak\,\mbox{MeV}}
\def\GeV{\nobreak\,\mbox{GeV}}

\def\pli{p^\prime}
\def\mli{{M^\prime}^2}
\def\mm{{M_m}^2}
\begin{document}

%\draft
%\preprint{\vbox{Submited to {\it Phys. Lett. \bf{B}}\hfill
%IFUSP/P-1344\\}}
%%%%%%%%% %%%%%%%%% %%%%%%%%% %%%%%%%%% %%%%%%%%% %%%%%%%%% %%
\title{\sc The  $J/\psi D D$ Vertex in QCD Sum Rules}
\author { R.D. Matheus, F.S. Navarra, 
M. Nielsen and R. Rodrigues da Silva\\
\vspace{0.3cm}
%{\it $^1$Instituto de F\'{\i}sica,  Universidade do Estado do Rio de Janeiro,}
% \\ 
%{\it Rua S\~ao Francisco Xavier 524, Maracan\~a,  20559-900, Rio de Janeiro, 
%RJ, Brazil}\\
%\vspace{0.3cm}
{\it Instituto de F\'{\i}sica, Universidade de S\~{a}o Paulo, } \\
{\it C.P. 66318,  05389-970 S\~{a}o Paulo, SP, Brazil}}
\maketitle
\vspace{1cm}

\begin{abstract}
The $J/\psi D D$ form factor is evaluated in a QCD sum rule calculation
for both $D$ and $J/\psi$ off-shell mesons. 
We study the double Borel sum rule for the three
point function of two pseudoscalar and one vector meson current.
We find that the momentum dependence of the form factors is  different
if the $D$ or the $J/\psi$ meson is off-shell, but they lead to the same 
coupling constant in the $J/\psi D D$ vertex. 
\\
PACS numbers 14.40.Lb,~~14.40.Nd,~~12.38.Lg,~~11.55.Hx
\\

\end{abstract}

\vspace{1cm}
%\newpage

The coupling constant of the strong interactions at the most fundamental level, i.e., 
of quarks and gluons, has been studied  both theoretically in lattice QCD calculations
and experimentally in $e^+ e^-$ colliders. Our knowledge on  $\alpha_s$ has 
significantly increased during the last decade \cite{mrst}. Unfortunately, these advances did
not bring a better understanding of the seemingly more accessible quantities: the
coupling constants bewteen hadrons. Indeed, on this next level of the strong 
interactions there has been little progress in what regards the determination of 
the ``hadron strong charges''. Even in the most studied case, the $N N \pi$ 
coupling, there is still some controversy. In the strange sector, the couplings between 
nucleons, hyperons and strange mesons have been strongly constrained by the use 
of SU(3) symmetry and intense phenomenological analyses of low energy hadronic
reactions
\cite{su3}. In the charm sector, neither it is  reasonable to use SU(4) symmetry since 
it is badly broken, nor experimental information is available. Of course, there are, 
here and there, exceptions as, for example, the $D^* D \pi$ coupling, which was recently 
determined by the CLEO collaboration from  $D^*$ decays \cite{cleo}.

In a curious development of the recent history of hadron physics, the couplings involving 
charm mesons became a very important building block in the construction of effective
lagrangian theories designed to describe low energy reactions such as, for example:
\beq
\pi + J/\psi \rightarrow D^* + \overline{D} 
\label{reac}
\eeq
Actually, apart from the couplings, more detailed knowledge is required, namely, 
we need to know the form factors at these effective vertices, as for example
$g_{D^* D \pi}(Q^2)$. These effective theories may be used to compute cross 
sections between light and charmed mesons, which, in turn may be used to 
understand charm hadron production (and suppression) in the nucleus-nucleus
collisions performed at RHIC \cite{mamu98,linko,lkz,ldk,haglin,haga,stt,osl,nnr,lkl}.

The work of calculating these form factors is rewarded with an interesting 
side product: information about the size of charm mesons. Of course, 
the size of a hadron depends on how
we ``look'' at it. The most extensively studied particle is the nucleon, 
which has been probed mainly by photons. In lower energy experiments, where 
also the four momentum transfer ($q^2$) is low,  it was possible to determine 
the electromagnetic form factor  (and the charge radius) of the nucleon. In 
higher energies experiments and very large values of $Q^2$ ($Q^2 = - q^2$) a very 
different picture of the nucleon emerged, in which it is made of pointlike particles, 
the quarks. From these  observations one may conclude that, when probing the 
nucleon, nearly on-shell photons ($q^2 \simeq 0$) recognize  sizes whereas 
highly off-shell photons ($q^2 << 0$) do not.  This statement is supported 
by the phenomenologically very successful vector meson dominance hypothesis, 
according to which real photons are with a large probability converted to 
vector mesons (which are extended objects) and then interact with the nucleon. 

A couple of years ago we started our program of computing the above mentioned 
quantities in the framework of QCD sum rules (QCDSR).
In \cite{nn} we calculated the coupling constant in the vertex 
$N D \Lambda_c$ with the $D$ meson off-shell. In \cite{bnn99} we did a 
similar calculation for strange coupling constants. In \cite{dnn} we extended the 
calculation performed in \cite{nn} and computed the $Q^2$ dependent form 
factors of the $N D \Lambda_c$ and $N D^* \Lambda_c$ vertices. We have also 
studied the  $D^* D \pi$, $B^* B \pi$  \cite{ddpi,ddpi2} and $D D \rho$ \cite{ddro} 
vertices.  One of the conclusions of these works 
is that when the off-shell particle in the vertex is heavy, the form 
factor tends to be  harder as a function of $Q^2$ , which 
means larger cut-off parameters and smaller associated sizes.

In the present work we will further investigate  form factors involving 
heavy mesons in order to extend our previous conclusions. We also want to 
better estimate the uncertainties in the procedure of determining coupling 
constants  with our techniques. For these 
purposes we consider the vertex $J/\psi D D$ and compute form factors and 
coupling constants for the cases where the $D$ is off-shell, the $J/\psi$ is 
off-shell and then compare the results. 

Following the QCDSR formalism  
described in our previous works we write the three-point function associated 
with a $J/\psi D D$ vertex, which is given by
\begin{equation}
\Gamma_\mu^{(D)}(p,\pli)=\int d^4x \, d^4y \, \langle 0|T\{j_D(x)
j_D^\dagger(y)j_\mu^\dagger(0)\}|0\rangle  
\, e^{i\pli.x} \, e^{-i(\pli-p).y}\; , 
\label{cord}
\end{equation}
for a $D$ off-shell meson, and by
\begin{equation}
\Gamma_\mu^{(J/\psi)}(p,\pli)=\int d^4x \, d^4y \, \langle 0|T\{j_D(x)
j_\mu^\dagger(y)j_D^\dagger(0)\}|0\rangle  
\, e^{i\pli.x} \, e^{-i(\pli-p).y}\; , 
\label{corro}
\end{equation}
for a $J/\psi$ off-shell meson,
where $j_D=i\bar{d}\gamma_5 c$, $j_D^\dagger=i\bar{c} \gamma_5d $ and
$ j_\mu^\dagger=\bar{c}\gamma_\mu c$ are the interpolating fields for the outgoing $D^+$, 
incoming $D^+$ and $J/\psi$ respectively with $d$ and $c$ being the down, and 
charm quark fields respectively.

The general expression for the vertex function, or correlator,  in Eqs.~(\ref{cord}) and
(\ref{corro}) has two independent structures. We can write 
$\Gamma_\mu$ in terms of the invariant
amplitudes associated with these two structures:
\beq
\Gamma_\mu(p,\pli)=\Gamma_1(p^2,{\pli}^2,q^2)p_\mu + \Gamma_2(p^2,{\pli}^2,q^2)
\pli_\mu\;
\eeq

As usual we shall write the correlators (\ref{cord}) and (\ref{corro}) in terms of hadron and 
quark-gluon degrees of freedom. These two representations, also called phenomenological and 
theoretical side, are then indentified one with the  other yelding a sum rule.  On the theoretical 
side we make an operator product expansion (OPE), keeping the first terms, which are represented 
in Fig.~1a and 1b for the correlators (\ref{cord}) and (\ref{corro})  respectively.  The diagram  of 
Fig.~1a and the first of Fig.~1b  may be evaluated giving the perturbative contributions. Applying 
Cutkosky's rule to these
contributions we can write  a double dispersion relation for each one of the invariant
amplitudes $\Gamma_i^{pert}\,(i=1,2)$, over the virtualities $p^2$ and ${\pli}^2$
holding $Q^2=-q^2$  ( with $q=\pli -p$) fixed:
\beq
\Gamma_i^{pert}(p^2,{\pli}^2,Q^2)=-{1\over4\pi^2}\int_{s_{min}}^\infty ds
\int_{m_c^2}^\infty du {\rho_i^{pert}(s,u,Q^2)\over(s-p^2)(u-{\pli}^2)}\;,
\label{dis}
\eeq
where $\rho_i^{pert}(s,u,Q^2)$ equals the double discontinuity of the amplitude
$\Gamma_i(p^2,{\pli}^2,Q^2)$ on the cuts $s_{min}\leq s\leq\infty$,
$m_c^2\leq u\leq\infty$, and where  $s_{min}=4 m_c^2$ in the case of the $D$ 
off-shell, where the dispersion relation is written in terms of the
two $D$ mesons' momenta, and  $s_{min}=m_c^2$ in the case of $J/\psi$ off-shell,
the dispersion relation being now written in terms of the $J/\psi$ and the
$D$ meson momenta. In terms of the $s,u$ and $t$ invariants, the double 
discontinuities are:

\beq
\rho_1^{pert(D)}=\frac{3}{4 \sqrt{\lambda}} \left[ 2 (u-m_c^2)+ (s-t-u+2 m_c^2) 
 (1 - \frac{(s-t-u-2 m_c^2) (s-t+u)}{\lambda})\right]
\label{d1}
\eeq
\beq
\rho_2^{pert(D)}= \frac{3 s}{2 \sqrt{\lambda}} \left[-1 + \frac{(s-t-u+2 m_c^2) (s-t-u-2 m_c^2)}
{\lambda} \right]
\label{d2}
\eeq
for a $D$ off-shell meson, with the integration limits for $u$:
%\beq
%(m_c^2+Q^2)(u-m_c^2)\geq sm_c^2\; ,
%\eeq

\beq
u_{max,min}=\frac{1}{2 m^2_c} \left[ -st+m_c^2 (s+2 t) \pm \sqrt{ (s-4 m_c^2)s (m_c^2-t)^2} \right]
\label{ulimD}
\eeq
and
\beq
\rho_1^{pert(J/\psi)} = -\frac{3}{\lambda^{3/2}} (s-t-u) (m_c^4 - s u)
\label{ro1}
\eeq

\beq
\rho_2^{pert(J/\psi)}  = \frac{3}{\lambda^{3/2}} (s+t-u) (m_c^4 - s u)
\label{ro2}
\eeq
for a $J/\psi$ off-shell meson, with the corresponding integration limits for $u$:
%\beq
%(s-m_c^2)(u-m_c^2)\geq Q^2m_c^2\; .\label{lro}
%\eeq

\beq
u_{max,min}=\frac{1}{2 m^2_c} \left[ -st+m_c^2 (2 s+t) \pm (s-m_c^2) \sqrt{ -t (4 m_c^2-t)} \right]
\label{ulimJ}
\eeq
with $ \lambda = \lambda(s,t,u)=s^2+t^2+u^2-2st-2su-2tu$ and $t=-Q^2$.   For this last limit we remark that  the condition
$u \geq t-m_c^2$ must be satisfied and  that, as it can be seen from the square root, in the timelike 
region, $t = q^2 >0$, we must have $ t  \geq 4 m_c^2$.

Since we are dealing with heavy quarks, we expect the perturbative contribution 
to be dominant on the OPE side.  However, for  the $J/\psi$ off-shell it turns out that 
the contribution of the quark condensate, shown in the second diagram of Fig.~1b,  is also important and 
must be included:
\beq
\Gamma_\mu^{<\bar{q}q>(J/\psi)}(p,\pli)={m_c \langle\bar{q}q\rangle\over 
(p^2-m_c^2)(
p^{'2}-m_c^2)} (p_\mu + p^{'}_\mu) \,\,=\,\,\Gamma^{<\bar{q}q>} \,\,(p_\mu + p^{'}_\mu)
\eeq

The phenomenological side of the vertex function is
obtained by considering the contribution of the $J/\psi$ and one $D$ meson, 
or the two $D$ mesons  states  to the matrix element
in Eqs.~(\ref{cord}) and (\ref{corro}) respectively.  In doing so, we introduce the 
decay constants $f_D$ and $f_{J/\psi}$, which are defined by the matrix elements
\beq
\langle 0|j_D|D\rangle={m_D^2f_D\over m_c}\;,
\label{fh}
\eeq
and
\beq
\langle J/\psi|j^\dagger_\mu|0\rangle=m_{J/\psi}f_{J/\psi}\epsilon^*_\mu
\; ,
\label{fd*}
\eeq
where $\epsilon_\mu$ is the polarization of the vector meson. We also make use of 
the matrix element:
\beq
\langle J/\psi(q) D(p')| D(p)\rangle=-g_{J/\psi D D}(Q^2) (p_{\mu} + p'_{\mu})\epsilon^\mu\;.
\label{con}
\eeq
Altogether these steps lead to:
\beqa
\Gamma_\mu^{phen(D)}(p,\pli)&=&-{m_D^2f_D\over m_c}m_{J/\psi} f_{\psi}
g_D(Q^2){1\over p^2-m_{J/\psi}^2}{1\over{\pli}^2-m_D^2}\times
\nonumber \\*[7.2pt]
&&\left(-2\pli_\mu+{m_D^2+m_{J/\psi}^2+Q^2\over m_{J/\psi}^2}p_\mu\right)
+ \mbox{higher resonances}\; ,
\label{phend}
\eeqa
\beqa
\Gamma_\mu^{phen(J/\psi)}(p,\pli)&=&-{m_D^4f_D^2\over m_c^2}
g_{J/\psi}(Q^2){1\over p^2-m_D^2}{1\over{\pli}^2-m_D^2}
%\nonumber \\*[7.2pt]
\left(\pli_\mu+p_\mu\right)
+ \mbox{higher resonances}\; ,
\label{phenro}
\eeqa
where
\beq
g_D(Q^2)={m_D^2f_D\over m_c}{g_{DDJ/\psi}^{(D)}(Q^2)\over Q^2+m_D^2}\;,
\label{gd}
\eeq
and
\beq
g_{J/\psi}(Q^2)= m_{J/\psi} f_{\psi} {g_{DDJ/\psi}^{(J/\psi)}(Q^2)\over Q^2+
m_{J/\psi}^2}\;, \label{gro}
\eeq
where $m_D$, $m_{J/\psi}$  are the masses 
of the mesons $D$ and $J/\psi$ respectively, and $g_{DDJ/\psi}^{(M)}(Q^2)$
is the form factor at the $DDJ/\psi$ vertex when the meson $M$ is off-shell.
The contribution of higher resonances in Eqs.~(\ref{phend})
and (\ref{phenro}) will be taken into account in the standard form of 
continuum contribution from the thresholds $s_0$ and $u_0$ \cite{io2}.

For consistency we use in our analysis the QCDSR expressions for 
the decay constants appearing in Eqs.~(\ref{phend}) and (\ref{phenro})
up to dimension four:
\beq
f_D^2={3m_c^2\over 8\pi^2m_D^4}\int_{m_c^2}^{u_0}du 
{(u-m_c^2)^2\over u}e^{(m_D^2-u)/\mm}\,-\, {m_c^3\over m_D^4}
\langle\bar{q}q\rangle e^{(m_D^2-m_c^2)/\mm}\; ,\label{fd}
\eeq
\beq
f_\psi^2={1\over4\pi^2}\int_{4m_c^2}^{r_0}dr~{(r+2m_c^2)
\sqrt{r-4m_c^2}\over r^{3/2}}e^{(m_{J/\psi}^2-r)/\mm}
\label{gJpsi}
\eeq
where $u_0 = (m_D + \Delta_u)^2 \GeV^2$,  $r_0 = (m_{J/\psi} + \Delta_r)^2 \GeV^2$ and
$\mm$ is the Borel mass used in the two-point functions given above.

We have omitted the  numerically insignificant contribution of the
gluon condensate. 

Eq.~(\ref{phenro}) shows that when the $J/\psi$ is off-shell, 
the sum rules in both structures give 
the same results.  On the other hand, according to Eq.~(\ref{phend}),  when the 
$D$ is off-shell the two structures give different results. In this last case
we will concentrate
in the $\pli_\mu$ structure, which we found to be the more stable one. The sum rules in 
this structure read:
\beq
\Gamma_2^{phen(D)} (p^2,{\pli}^2,q^2) =  \Gamma_2^{pert(D)} (p^2,{\pli}^2,q^2) 
\label{sumruled}
\eeq
for a $D$ off-shell and 
\beq
\Gamma_2^{phen(J/\psi)} (p^2,{\pli}^2,q^2) =  \Gamma_2^{pert(J/\psi)} (p^2,{\pli}^2,q^2) + 
\Gamma^{<\bar{q}q>} (p^2,{\pli}^2,q^2) 
\label{sumrulepsi}
\eeq
for a $J/\psi$ off-shell.

Inserting in these equations the corresponding expressions for the 
perturbative, quark condensate and phenomenological terms and 
performing  a double Borel transformation \cite{io2} in both variables
$P^2=-p^2\rightarrow M^2$ and ${P^\prime}^2=-{\pli}^2\rightarrow \mli$ we obtain
the final expressions for the sum rules. In order to allow for different  values 
of $M^2$ and $\mli$ we take them proportional to the respective meson masses, 
which leads us to study the sum rules as a function of $M^2$ at a fixed ratio:
\beq
\frac{M^2}{\mli} = \frac{m_{J/\psi}^2}{m_D^2}
\label{relmass}
\eeq
for (\ref{sumruled}) and ${M^2}={\mli}$ for (\ref{sumrulepsi}).

In refs.~\cite{BBG93,ra} it was found that relating 
the Borel parameters in the two- $(\mm)$ and three-point functions $(M^2)$
as
\begin{equation}
         2M_{m}^2=M^2 ~ ,
\label{borel}
\end{equation}
is a crucial
ingredient for the incorporation of the HQET symmetries, and leads
to a considerable reduction of the sensitivity to input parameters, 
such as continuum thresholds $s_0$ and $u_0$, and to radiative 
corrections. Therefore, in this work we will use Eq.~(\ref{borel}) to relate
the Borel masses.

The parameter values used in all calculations are $m_q=(m_u+m_d)/2=7\,\MeV$, 
$m_c=1.3\,\GeV$, $m_D=1.87\,\GeV$, $m_{J/\psi}=3.1\,\GeV$, 
$\langle\overline{q}q\rangle\,=\,-(0.23)^3\,\GeV^3$. 
 We parametrize the continuum thresholds as
\beq
s_0=(m_{M}+\Delta_s)^2\;,\label{s0}
\eeq
where $m_M=m_D(m_{J/\psi})$ for the case that the $J/\psi(D)$ meson is off-shell,
and
\beq
u_0=(m_D+\Delta_u)^2\;.\label{u0}
\eeq

Using  $\Delta_s = \Delta_u = \Delta_r \sim0.5\GeV$ for the continuum thresholds and 
fixing $Q^2$ we found good stability of the sum rule for $g_{DDJ/\psi}^{(D)} (Q^2)$ as 
a function of $M^2$ in the interval $ 4 < M^2 < 16 \GeV^2$  and also for 
$g_{DDJ/\psi}^{(J/\psi)}(Q^2)$ in the interval
$ 2 < M^2 < 10 \GeV^2$.

Fixing $M^2 =  11 \,\GeV^2$ we calculate 
the momentum dependence of the form factor $g_{DDJ/\psi}^{(D)}(Q^2)$
in the interval $- 0.5 \leq Q^2
\leq 5.0 \, GeV^2 $, where we expect the sum rules to be valid (since in this 
case the cut in the $t$ channel starts at $t\sim m_c^2$ and thus the Euclidian
region stretches up to that threshold). Our numerical calculations can be 
well reproduced by the gaussian parametrization:
\beq
g_{DDJ/\psi}^{(D)}(Q^2)= 16.4 \,\, e^{-[ \frac{(Q^2+16.2)^2}{228} ]}
\label{exp}
\eeq

We stress here that it was not possible to fit our results with a monopole form!

As in ref.~\cite{lkz},  we define the coupling constant as  
the value of the form factor at $Q^2=-m_M^2$, where $m_M$ is the 
mass of the off-shell meson. In the case of an off-shell $D$, this leads to
\beq
g_{DDJ/\psi}^{(D)} = 8.05
\label{gd1}
\eeq

Along the same lines we choose  $M^2 =  8 \,\GeV^2$ and calculate 
the momentum dependence of the form factor $g_{DDJ/\psi}^{(J/\psi)}(Q^2)$. 
The numerical results can be fitted by the monopole parametrization:

\beq
g_{DDJ/\psi}^{(J/\psi)}(Q^2)=  \frac{1069.76}{Q^2+143.18}
\label{mo}
\eeq
which, at the $J/\psi$ pole leads to:
\beq
g_{DDJ/\psi}^{(J/\psi)} = 7.98
\label{gpsi}
\eeq

All our numerical results and parametrizations are shown in Fig.~2. The circles 
and squares correspond to the numerical results for the $D$ and $J/\psi$ 
off-shell respectively. These points are fitted by  the dashed (Eq. (\ref{exp})) and solid lines 
(Eq. (\ref{mo})) respectively and extrapolated to the meson poles which are represented by 
the triangles, which give the numbers quoted in (\ref{gd1}) and (\ref{gpsi}). 

A closer look into Fig.~2 reveals that, whereas the 
circles are well fitted by the dashed line, this is not the case of the squares, which 
are not very  accurately described by the solid line. Due to this reason and also because the
$J/\psi$ pole is farther away, the estimate (\ref{gpsi}) is much less reliable than 
(\ref{gd1}). In fact, to reduce the uncertainty in the extrapolation of the $J/\psi$ off-shell  
form factor, we have to impose the condition that both coupling constants  $g_{DDJ/\psi}^{(D)}$
and  $g_{DDJ/\psi}^{(J/\psi)}$ must coincide. We therefore use the former as a guide in the 
fitting procedure (and subsequent extrapolation) leading to the latter. This condition also 
imposes a  severe constraint in the  minimum $Q^2$  used in the calculation. We have used 
the interval $5.5 < Q^2 <  9.5 \GeV^2$. Including smaller values of $Q^2$ in the calculation 
would increase the curvature of the solid line, which, when extrapolated to lower $Q^2$ values, 
would give coupling constants very different from  $g_{DDJ/\psi}^{(D)}$. In other words, it 
would become impossible to obtain the two triangles at the same height.

The star in Fig.~2 indicates the $J/\psi$ off-shell form factor taken at $Q^2=0$. 
As it can be immediately obtained  from (\ref{mo}), 
$g_{DDJ/\psi}^{(J/\psi)} = 7.47$ at $Q^2=0$. This coincides with the estimate
made in Ref. \cite{mamu98} using the vector meson dominance model.

The information presented in Fig.~2 may now be crossed with those  obtained in \cite{ddro}, 
about the size of the $D$ meson. In Fig.~3 we compare $g_{DDJ/\psi}^{(J/\psi)}(Q^2)$, 
given by (\ref{mo}) with $g_{DD\rho}^{(\rho)}(Q^2)$, calculated in \cite{ddro} and parametrized by:
\beq
g_{DD\rho}^{(\rho)}(Q^2)= 2.53 \,\, e^{-{Q^2\over 0.98}}
\label{expro}
\eeq
The $DDJ/\psi$ form factor is represented by a solid line whereas the $DD\rho$ one is shown with a
dashed line. 

On a qualitative level, the comparison between solid and dashed lines both in Fig.~2 and Fig.~3 shows  
that the form factor is harder if the off-shell meson is heavier. In particular, remembering that  there 
is an overlap between the photon and the vector mesons $\rho$ and $J/\psi$, we check in Fig.~3 an  
empirical formula regarding the resolving power of the photon, used some time ago by experimentalists.   
HERA data on electron-proton reactions could  be well understood 
introducing a ``transverse radius of the photon'', parametrized as \cite{cart}:
\beq
 r_{\gamma} \simeq 1 / \sqrt{Q^2 + m^2} 
\label{racar}
\eeq
where  $m$ is the mass of the vector meson considered. This 
empirical formula tells us that for $Q^2 \rightarrow \infty$ the photon is 
pointlike  and ``resolves'' the nucleon target, i.e., identifies its 
pointlike constituents and does not ``see''  the size of the nucleon.
Moreover this formula indicates that for $Q^2 \simeq 0$ and for light mesons
(like the $\rho^0$) the photon has appreciable transverse radius  and therefore 
also identifies the global nucleon extension. Finally, in the above formula we 
may have a heavy vector meson ($J/\psi$ or $\Upsilon$) which will, either 
real or virtual, resolve the nucleon into pointlike constituents. 
This conjecture can be applied to a $D$ target probed by a $J/\psi$and 
seems to be supported by Fig.~3 when we consider the vicinity of $Q^2 \simeq 0$. There, 
we see (again) that the slope of the dashed line is stronger than that of the 
solid line.  This means that also around the ``real photon'' region, the $J/\psi$ 
resolves smaller scales then the $\rho$.

In Fig.~4 we compare $g_{DDJ/\psi}^{(D)}(Q^2)$, given by (\ref{exp}) (solid line), with   
$g_{DD\rho}^{(D)}(Q^2)$ (dashed line), calculated in \cite{ddro} and parametrized by:

\beq
g_{DD\rho}^{(D)}(Q^2)=  \frac{37.5}{Q^2+12.12}
\label{mod}
\eeq
We can clearly observe that the $D + \rho \rightarrow D$ transition has a much harder form factor 
than the  $D + J/\psi \rightarrow D$ one.  

In the limiting case $r_{J/\psi} << r_{D} << r_{\rho} $  these results 
can be understood in a simple picture: when the $D$ hits the smaller $J/\psi$ it can 
``see'' a size, it can measure it! On the other hand, when  the $D$ hits the much larger $\rho$ meson, 
it does not see any size, in the same way as large $Q^2$ photons (in DIS measurements) do not see any 
size in the proton. Rather, they will interact with pointlike partons! In real life the mentioned radii 
are not so different from each other and therefore the differences in the curves in Fig.~4 are not so 
pronounced.

Our calculations contain uncertainties in the QCD parameters (masses and vacuum condensates), in
the  OPE (because we neglect higher order operators), in the model of the continuum (the uncertainty 
in the values of $s_0$ and $u_0$) and the systematic error in the extrapolation procedure to obtain 
the couplings. These errors are present in most of the QCDSR calculations and are to a certain 
extent unavoidable. Due to them most of QCDSR results are plagued by a $20  \%$ error. Therefore
our numbers for the couplings have this same uncertainty. On the other hand, the curves shown in the 
figures are so dramatically different that they confirm beyond any doubt our previous suspiction 
\cite{dnn,ddpi,ddro} regarding  transition vertices and, most of all, as one can see clearly in Fig.~3,
they  show in the case of ``charge form factors'' (where the same meson goes in and out the vertex), 
that the $D$ meson ``seen'' by a $\rho$ is much larger than when it is probed by the $J/\psi$.

Our program will continue and studies of other vertices are  
in progress.

\vspace{1cm}
 
\underline{Acknowledgements}: 
This work has been supported by CNPq and FAPESP. We are indebted to prof. E. Shuryak for
fruitful discussions.

\begin{figure} \label{fig0}
%\leavevmode
\begin{center}
\vskip -1cm
\epsfysize=9.0cm
\epsffile{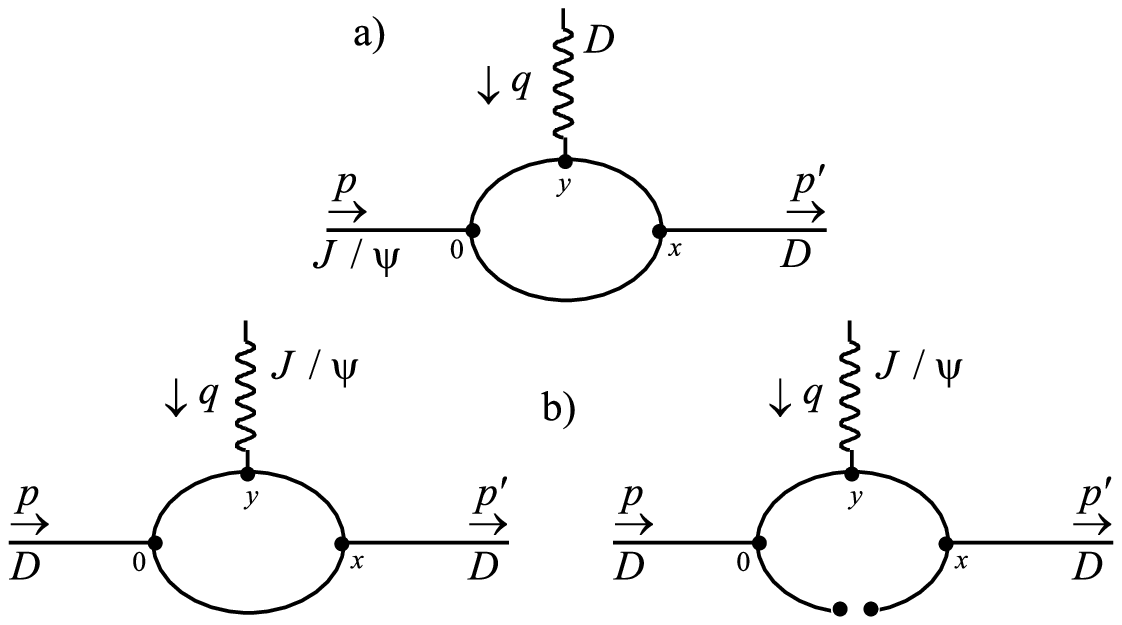}
\caption{ a) diagrams that contribute to  $g_{DDJ/\psi}^{(D)}(Q^2)$ 
b) diagrams that contribute to  $g_{DDJ/\psi}^{(J/\psi)}(Q^2)$}
\end{center}
\end{figure}

\begin{figure} \label{fig1}
%\leavevmode
\begin{center}
\vskip -1cm
\epsfysize=9.0cm
\epsffile{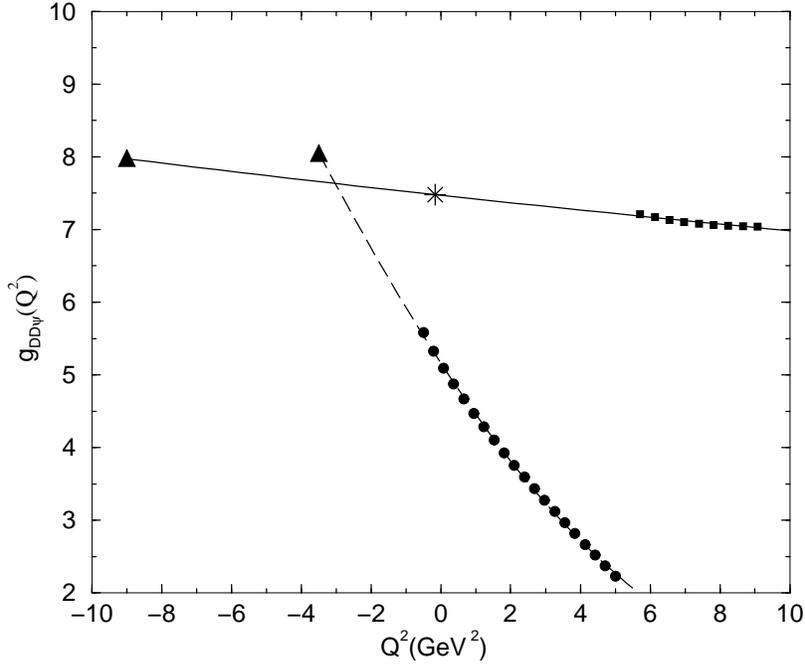}
\caption{Momentum dependence of the $DDJ\psi$ form factor. 
Circles and squares represent our numerical calculations for the $D$ and $J/\psi$ off-shell
respectively.  The dashed and solid lines give the parametrization of the QCDSR results through 
Eq.~(\protect\ref{exp}) for the circles and Eq.~(\protect\ref{mo}) for the squares. The triangles 
give the form factors at the poles of the particles (which we indentify with the coupling constant).
The star shows the form factor at $Q^2=0$.}
\end{center}
\end{figure}

\begin{figure} \label{fig2}
%\leavevmode
\begin{center}
\vskip -1cm
\epsfysize=9.0cm
\epsffile{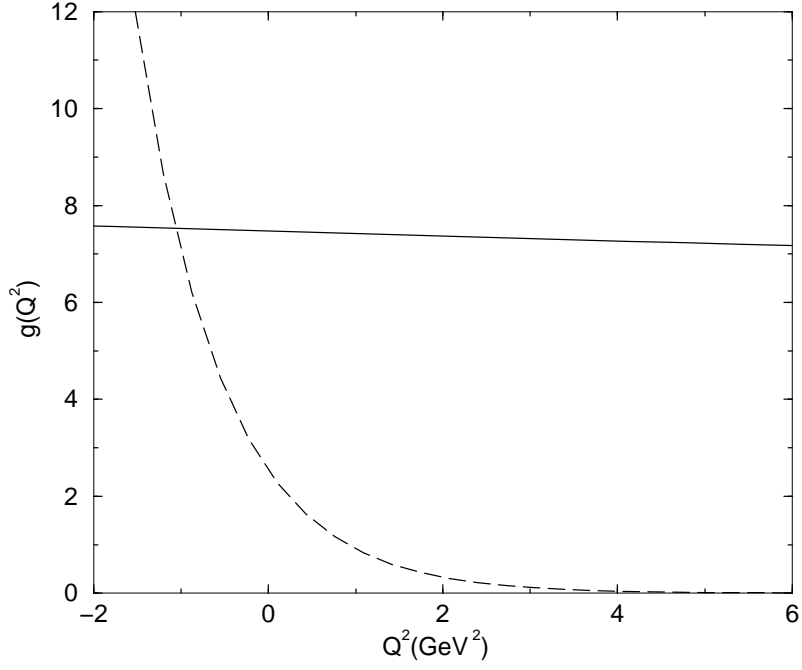}
\caption{Momentum dependence of the $DDJ\psi$ (solid line) and $DD\rho$ 
(dashed line) form factors. In both cases the vector mesons are off-shell.}
\end{center}
\end{figure}

\begin{figure} \label{fig3}
%\leavevmode
\begin{center}
\vskip -1cm
\epsfysize=9.0cm
\epsffile{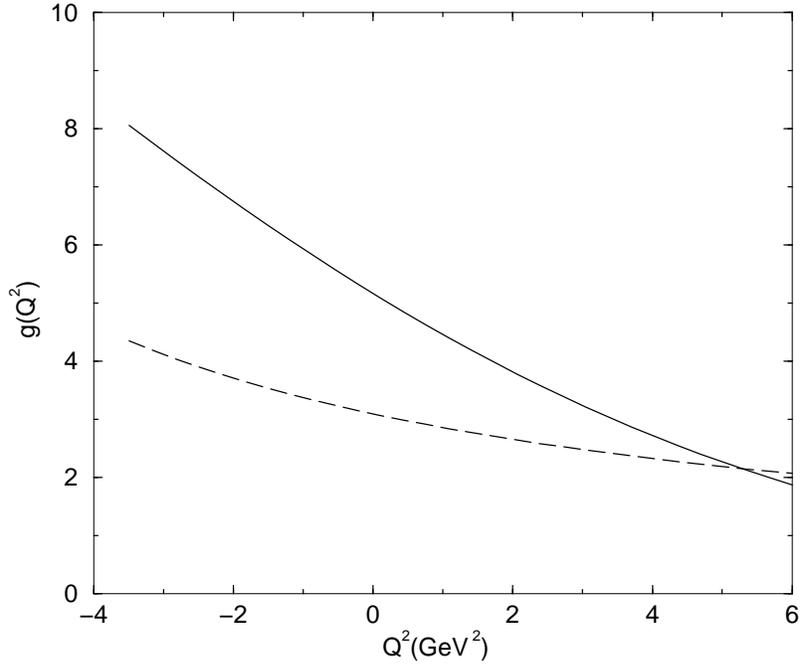}
\caption{Momentum dependence of the $DDJ\psi$ (solid line) and $DD\rho$ 
(dashed line) form factors. In both cases the $D$ meson is off-shell.}
\end{center}
\end{figure}

\end{document}